\input{aipcheck}

\documentclass[
    ,draft            % use draft while you are working on the paper
    ,numberedheadings % uncomment this option for numbered sections
  ]
  {aipproc}

\layoutstyle{6x9}

\begin{document}

\title{Surprising Aspects of Fluctuating ``Pulled'' Fronts}

\author{Debabrata Panja}
{address={Instituut-Lorentz, Universiteit Leiden, Postbus 9506, 2300
RA Leiden, The Netherlands\footnote{Present Address: Institute for
Theoretical Physics, Universiteit van Amsterdam, Valckenierstraat 65,
1018 XE Amsterdam, The Netherlands}}}

\begin{abstract}
\noindent Recently it has been shown that when an equation that allows
so-called pulled fronts in the mean-field limit is modelled with a
stochastic model with a finite number $N$ of particles per correlation
volume, the convergence to the speed $v^*$ for $N \to \infty$ is
extremely slow --- going only as $\ln^{-2}N$. However, this convergence
is seen only for very high values of $N$, while there can be
significant deviations from it when $N$ is not too large. Pulled
fronts are fronts that propagate into an unstable state, and the
asymptotic front speed is equal to the linear spreading speed $v^*$ of
infinitesimal perturbations around the unstable state. In this paper,
we consider front propagation in a simple stochastic lattice
model. The microscopic picture of the front
dynamics shows that for the description of the far tip of the front,
one has to abandon the idea of a uniformly translating front
solution. The lattice and finite particle effects lead to a
``halt-and-go'' type dynamics at the far tip of the front, while the
average front behind it ``crosses over'' to a uniformly translating
solution. In this formulation, the effect of stochasticity on the
asymptotic front speed is coded in the probability distribution of the
times required for the advancement of the ``foremost occupied lattice
site''. These probability distributions are obtained by matching the
solution of the far tip with the uniformly translating solution behind
in a mean-field type approximation, and the results for the
probability distributions compare well to the results of stochastic
numerical simulations. This approach allows one to deal with much
smaller values of $N$ than it is required to have the $\ln^{-2}N$
asymptotics to be valid.
\end{abstract}


\maketitle

\section{The Basics of Front Propagation: Pulled Fronts}

In pattern forming systems quite often situations occur where  patches
of different bulk phases occur which are separated by fronts or
interfaces. In such cases, the relevant dynamics is usually dominated
by the dynamics of these fronts. When the interface separates two
thermodynamically stable phases, as in crystal-melt interfacial growth
problems, the width of the interfacial zone is usually of atomic
dimensions. For such systems, one often has to resort to a moving
boundary description in which the boundary conditions at the interface
are determined phenomenologically or by microscopic considerations. A
question that naturally arises for such interfaces is the influence of
stochastic fluctuations on the motion and scaling properties of such
interfaces.

At the other extreme is a class of fronts which arise in systems that
form patterns, and in which the occurrence of fronts or transition
zones is fundamentally related to their nonequilibrium nature, as they
do not connect two thermodynamic equilibrium phases which are
separated by a first order phase transition.  In such cases --- for
example, chemical fronts \cite{meron}, the temperature and density
transition zones in thermal plumes \cite{tabeling}, the domain walls
separating domains of different orientation in  in rotating
Rayleigh-B\'enard convection \cite{tucross}, or streamer fronts in
discharges \cite{streamers} --- the fronts are relatively wide and
therefore described by the same continuum equations that describe
nonequilibrium bulk patterns. The lore in nonequilibrium pattern
formation is that when the relevant length scales are large, (thermal)
fluctuation effects are relatively small \cite{ch}. For this reason,
the dynamics of many pattern forming systems can be understood in
terms of the deterministic dynamics of the basic patterns and coherent
structures. For fronts, the first questions to study are  therefore
properties like existence and  speed of propagation of the front
solutions of the deterministic equations.
\begin{figure}
%\begin{center}
\includegraphics[width=0.9\linewidth]{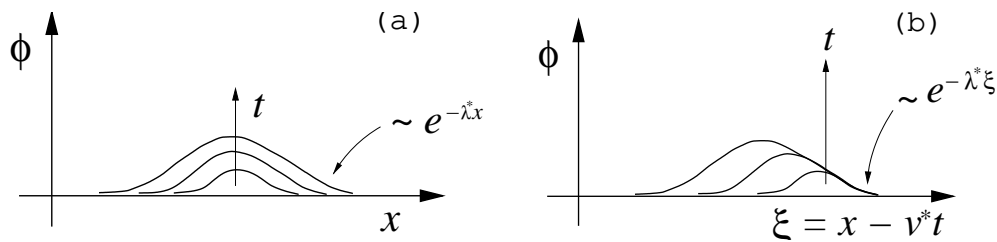}
\caption{Illustration of $v^*$ as the linear spreading speed of
infinitesimal perturbations around the unstable state. \label{fig1}
\vspace{-4mm}}
%\end{center}
\end{figure}
A class of fronts, for which these questions can be answered
theoretically, are the so-called pulled fronts. Pulled fronts are the
fronts that propagate into a linearly unstable state, and whose
asymptotic front speed is equal to the linear spreading speed $v^*$ of
infinitesimal perturbations around the unstable state
\cite{bj,vs2,ebert}. The name pulled front refers to the picture that
in the leading edge of these fronts, the perturbation around the
unstable state grows and spreads with speed $v^*$, while the rest of
the front gets ``pulled along'' by the leading edge. The intuitive
idea is captured in Fig. \ref{fig1}, where we consider the
(exaggerated) growth and spreading of an infinitesimal perturbation
around the linearly unstable state. Three snapshots of this
perturbation, taken at three different time instants in the laboratory
frame are plotted in Fig. \ref{fig1}(a). The initial perturbation is
chosen in such a way that it decays as $\exp[-\lambda^*x]$ for
$x\rightarrow\infty$, where $\lambda^*$ is the exponent associated
with $v^*$. The special status that the quantities $\lambda^*$ and
$v^*$ have in the growth and spreading of this perturbation is coded
in the fact that the $\exp[-\lambda^*x]$ decay of the perturbation for
$x\rightarrow\infty$ remains preserved at {\it all\/} stages of its
development. This fact is further illustrated in Fig. \ref{fig1}(b),
where the same three snapshots [as in Fig. \ref{fig1}(a)] have been
plotted in the comoving frame, moving with speed $v^*$ w.r.t. the
laboratory frame. That the notion of the leading edge ahead of the
front grows and spreads with speed $v^*$ and thereby pulls the rest of
the front along with it is not merely an intuitive picture but can be
turned into a mathematically precise analysis is illustrated by the
recent derivation of exact results for the general  power law
convergence of the front speed to the asymptotic value $v^*$
\cite{ebert}. On the other hand, fronts which propagate into a
linearly unstable state and whose asymptotic speed is $ >v^*$ are
referred to as pushed, as it is the nonlinear growth in the region
behind the leading edge that pushes their front speed to higher
values.  If the state is not linearly unstable, then $v^*$ is
trivially zero; in such cases front propagation is always
dominated by the nonlinear growth in the front region itself, and
hence fronts in this case are in a sense pushed too. To obtain the
asymptotic speed of a pushed front, one has to solve the full
nonlinear equation; in general it is not possible to do so except for
a special set of parameter values.
\begin{figure}
%\begin{center}
\includegraphics[width=0.5\linewidth]{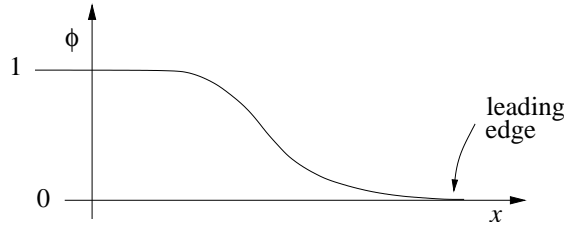}
\caption{An instantaneous snapshot of the front configuration of
Eq. (\ref{e1}). \label{fig2}\vspace{-4mm}}
%\end{center}
\end{figure}
To provide a more quantitative flavour of how the growth and spreading
of infinitesimal perturbation around the unstable state play a very
important role for pulled fronts, let us now consider and examine an
example of a deterministic equation that admits pulled fronts. The
equation that we choose for this purpose is the so-called Fisher
equation, which was at first used to model the spreading of
advantageous genes in a population \cite{fisher}. In this model, the
density of the advantageous genes is denoted by $\phi(x,t)$, whose
dynamics is described by means of the equation
\begin{eqnarray}
\frac{\partial\phi}{\partial t}\,=\,D\,\frac{\partial^2\phi}{\partial
t^2}\,+\,\phi\,-\,\phi^n.\quad\quad n>1,\quad\mbox{for example 2 or 3}
\label{e1}
\end{eqnarray}
Equation (\ref{e1}) has two stationary states, of which $\phi(x,t)=0$
is (linearly) unstable and $\phi(x,t)=1$ is stable. Therefore, if the
system is prepared in a way such that these two states coexist in a
certain region of space, then the stable state invades the unstable
one and propagates into it. An instantaneous configuration of the
resulting front is shown in Fig. \ref{fig2}.

\begin{figure}
%\begin{center}
\includegraphics[width=0.3\linewidth]{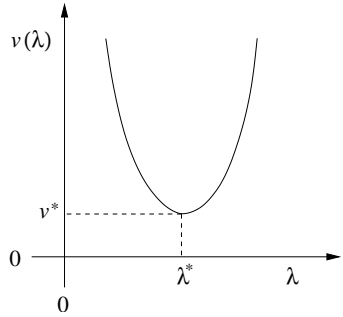}
\caption{The dispersion relation (\ref{e3}) is schematically shown
above. \label{fig3}\vspace{-4mm}}
%\end{center}
\end{figure}

To obtain the front solution admitted by Eq. (\ref{e1}), we rewrite it
in a frame that moves w.r.t. the laboratory frame at a constant speed
$v$, and look for a stationary solution of $\phi$ in this comoving
frame. In terms of the comoving co-ordinate $\xi=x-vt$, Eq. (\ref{e1})
can be simply rewritten by means of a change of variables from $(x,t)$
to $(\xi,t)$, as
\begin{eqnarray}
\frac{\partial\phi}{\partial
t}\,-\,v\,\frac{\partial\phi}{\partial\xi}\,=\,D\,\frac{\partial^2\phi}{\partial
\xi^2}\,+\,\phi\,-\,\phi^n.
\label{e2}
\end{eqnarray}
The crucial relevance of the growth and spreading of infinitesimal
perturbations enters naturally in this front solution, as the
propagating infinitesimal perturbations around the unstable state in
the leading edge ahead of the front sets on the instability making way
for further growth. At the leading edge of the front, the
$\phi$-values are very close to the unstable state value, i.e.,
$\phi\ll1$, and one can neglect the nonlinear term $\phi^n$ compared
to $\phi$ in Eq. (\ref{e2}). The stationary solution of the resulting
linear equation can then be solved by using
$\phi\sim\exp[-\lambda\xi]$, yielding the following relation between
$v$ and $\lambda$:
\begin{eqnarray}
v(\lambda)\,=\,D\,\lambda\,+\,\frac{1}{\lambda}\,.
\label{e3}
\end{eqnarray}
The curve for the dispersion relation between $v(\lambda)$ and
$\lambda$ is schematically shown in Fig. \ref{fig3}. It has a minimum
at $\lambda^*$, and $v^*=v(\lambda^*)=2\sqrt{D}$. The actual
dispersion relation depends on the model that one studies, but the
U-shape is a characteristic of fronts that propagate into a linearly
unstable state.  Although Eq. (\ref{e3}) indicates that Eq. (\ref{e1})
has a front solution for all values of $\lambda$ (and correspondingly
all possible front speeds), from which it might a priori seem that the
quantities $\lambda^*$ and $v^*$ are not special in any way, the
actual selection of the asymptotic front speed is obtained only after
a proper stability analysis of the front profile in the comoving
frame. Such a stability analysis yields the result that with the
initial condition that $\phi(x,t)|_{t=0}$ that decays faster than
$\exp[-\lambda^*x]$ for $x\rightarrow\infty$, the front speed
converges uniformly\footnote{Uniform convergence means that the
convergence behaviour (\ref{e4}) of the front speed is the same
irrespective of the value of $\phi$ at which the speed is being
measured.} to $v^*$ as \cite{ebert,bramson}
\begin{eqnarray}
v(t)\,=\,v^*\,-\,\frac{3}{2\lambda^* t}\,+\,{\cal O}(t^{-3/2})\,,
\label{e4}
\end{eqnarray}
as the front shape relaxes to its asymptotic steady state
configuration $\phi^*(x-v^*t)$.

\section{Fluctuating ``Pulled'' Fronts: A Different Paradigm}

From the above perspective, it is maybe less of a surprise that the
detailed questions concerning the stochastic properties of inherently
nonequilibrium fronts have been addressed, to some extent, only
relatively recently
\cite{breuer,lemar,armeroprl,bd,kns,levine,armero,riordan}, and that
it has taken a while for researchers to become fully aware of the fact
that the so-called pulled fronts \cite{bj,vs2,ebert,dee} that
propagate into an unstable state, do  {\em not} fit into the common
mold: they have anomalous sensitivity to particle effects
\cite{bd,kns,levine,ramses,pvs2}, and have been argued to display
uncommon scaling behavior \cite{riordan,rocco,tripathy1,tripathy2,moro}.

All these effects have one origin in common, namely the fact that the
dynamics of pulled fronts, by its very nature, is not determined by
the nonlinear front region itself, but by the region {\em at the
leading edge of the front},  where deviations from the unstable state
are small. To a large  degree, this semi-infinite region alone
determines the  universal relaxation of the  speed of a deterministic
pulled front to its asymptotic value \cite{vs2,ebert,bd}, as well as
the anomalous scaling behavior of stochastic fronts
\cite{rocco,tripathy1,tripathy2} in continuum equations with
multiplicative noise. The crucial importance of the region, where the
deviations from the unstable state are small, also implies that if one
builds a lattice model version of a front propagating from a stable
into a linearly unstable state, the front speed is surprisingly
sensitive to the dynamics of the tip (the far end) of the front where
only one or a few particles per lattice site are present. This is the
main subject of this paper, and we will demonstrate the effect of
discreteness on front propagation by means of considering a
discrete particle model of Fisher equation (\ref{e1}).

Following Refs. \cite{breuer,pvs}, the discrete particle model of
Fisher equation (\ref{e1}) that we will consider in this paper is that
of the reaction-diffusion system X $\leftrightarrow$ 2X on a
lattice. In this system, an X particle on any lattice site can diffuse
to one of its nearest neighbour lattice sites with a diffusion rate
$D$. The rate of the forward reaction X $\rightarrow$ 2X is $N/2$,
while the rate of the backward reaction is normalized to unity. A
forward reaction on a lattice site $k$ increases the number of X
particles on lattice site $k$ by one, and a backward 2X $\rightarrow$
X reaction on lattice site $k$ reduces the number of X particles on
lattice site $k$ by one. For large $N$ values, the microscopic
interaction indicates that an initial conglomeration of X particles
grows both in size and spread. The microscopic rules also dictate that
the growth saturates when the number of particles on a lattice site
reaches approximately $N$ --- at that stage, on average, the amount of
new X particles generated due to the forward reaction equals the
amount of X particles annihilated due to the backward reaction.

The deterministic mean-field limit of the this reaction-diffusion
system yields the equation \cite{breuer,pvs}
\begin{eqnarray}
\frac{\partial\phi_k}{\partial
t}\,=\,D\left[\phi_{k+1}\,+\,\phi_{k-1}\,-2\phi_k\right]\,+\,\phi_k\,-\,\phi_k^2\,.
\label{e5}
\end{eqnarray}
Here, $\phi_k=\langle N_k\rangle/N$, where $\langle N_k\rangle$ is the
conditionally (ensemble) averaged\footnote{In view of the stochasticity
in the system, one has to be careful about taking averages. A simple
ensemble averaging does not yield a steady front shape due to
diffusive wandering of individual front realizations. This notion is
best understood by means of Fig. 5 of Ref. \cite{breuer}, which shows
the ensemble average front shape by means of hashed lines. One needs
to filter out the diffusive wandering of fronts before taking the
average, and this is what is meant by conditional (ensemble)
averaging. For a detailed discussion, see Sec. III.B of
Ref. \cite{pvs}} number of X particles on the $k$-th lattice
site. Equation (\ref{e5}) is the lattice version of Eq. (\ref{e1}),
and it admits a pulled front solution, which can be obtained by using
the uniformly translating solution $\phi_k(t)\equiv\phi(k-vt)$. From
the analysis and discussion of the previous section, therefore, one
can expect that the asymptotic front speed is $v^*$, where the
corresponding front solution $\phi^*(k-v^*t)$ behaves
$\sim\exp[-\lambda^*\xi]$ for $\xi\rightarrow\infty$. Here
$\xi=k-v^*t$ and $\lambda^*$ and $v^*$ correspond to the minimum of
the dispersion relation $v(\lambda)$ vs. $\lambda$ for Eq. (\ref{e5})
(similar to Fig. \ref{fig3}). In actuality, however, the asymptotic
speed turns out to be $<v^*$. This was first observed numerically in
Ref. \cite{breuer}.

\begin{figure}
%\begin{center}
\includegraphics[width=0.65\linewidth]{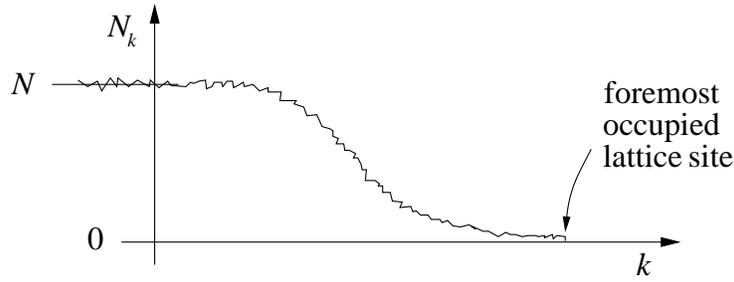}
\caption{An instantaneous snapshot of a front realization (travelling
from the left to the right) for the reaction diffusion system X
$\leftrightarrow$ 2X, at time $t\rightarrow\infty$. Note the
existence of the foremost occupied lattice site, on the right of which
no lattice site has been previously occupied. \label{fig4}\vspace{-4mm}}
%\end{center}
\end{figure}

The reason behind having an asymptotic speed $<v^*$ stems from
the discreteness of the particles and the lattice effects, and is
understood quite simply when one takes an instantaneous snapshot of
the resulting front, moving from the left to the right, for one single
realization at time $t\rightarrow\infty$\footnote{The limit
$t\rightarrow\infty$ is necessary to ensure that the front relaxes to
a steady shape in an average sense, so that there are no residual
large scale transient effects.}. Such a snapshot is shown in
Fig. \ref{fig4}. The important aspect to notice in Fig. \ref{fig4} is
that unlike the semi-infinite region that the leading edge occupies
for a deterministic front (as shown in Fig. \ref{fig2}), there exists
a foremost occupied lattice site (f.o.l.s.) in this snapshot, on the
right of which no lattice site has ever been occupied before. In the
language of the mathematicians, this means that the actual front
region has a compact support.
The crucial role played by the discreteness effects of the particles
and the lattice is reflected in the mechanism of the front propagation
at the f.o.l.s., where the mechanism is {\it not\/} that of a uniform
translation, but instead, is of {\it halt-and-go, which in this
reaction-diffusion model is diffusion dominated\/}, as we demonstrate
below. First of all, it is clear from the very nature of the
reaction-diffusion equation that unless there is at least one X
particle on a lattice site, the growth in the number of particles on
that lattice site does not take place. This means that the position of
the f.o.l.s. moves towards the right {\it only\/} when a particle from
the f.o.l.s. makes a diffusive hop towards the right. The movement of
the f.o.l.s. is therefore diffusion dominated, as opposed to being
also driven by growth; i.e., the phenomenon of leading edge ahead of
the front setting on the instability making way for further growth
that takes place for a deterministic mean field equation of a pulled
front, does not take place for the corresponding discrete particle
realization --- this makes the fronts made of discrete particles on a
lattice ``pulled'' as opposed to being pulled. Secondly, these
diffusive hops are not continuous in time, i.e., there is a finite
time difference between any two successive forward movements of the
f.o.l.s. (this is what is meant by halt-and-go; the f.o.l.s. halts for
the times between the successive hops), and these time differences are
stochastic in nature. We stress here that the halt-and-go mechanism of
the dynamics of the f.o.l.s. is a generic consequence of having
discrete particles on a discrete lattice; in the present model that we
consider, the movement is diffusion dominated, but there can be other
models where the f.o.l.s. moves by means of some other mechanism (see
for example, clock model \cite{ramses}).

The subtlety that the movement of the f.o.l.s. is not driven by growth
for ``pulled'' fronts made of discrete particles on a lattice
was first realized by Brunet and Derrida \cite{bd}, who chose to
implement it by having a growth cutoff at a $\phi_k$ value $1/N$ in
the deterministic mean-field Eq. (\ref{e5}). The choice of $1/N$ was
motivated by the fact that the number of particles on the f.o.l.s. is
${\cal O}(1)$, which corresponds to a field value $\phi_k\sim1/N$. The
uniformly translating solution of the corresponding mean-field front
equation for the reaction-diffusion process X $\leftrightarrow$ 2X 
\begin{eqnarray}
\frac{\partial\phi_k}{\partial
t}\,=\,D\left[\phi_{k+1}\,+\,\phi_{k-1}\,-2\phi_k\right]\,+\,\left[\phi_k\,-\,\phi_k^2\right]\,\Theta\left(\phi_k\,-\,\frac{1}{N}\right)\,,
\label{e6}
\end{eqnarray}
where the $\Theta$ term is the heavyside theta function\footnote{The
``pulled'' nature can be made mathematically very precise for
Eq. (\ref{e6}), see for example Ref. \cite{pvs3}.}, yields an
asymptotic speed $v_{cutoff}$, given by
\begin{eqnarray}
v_{cutoff}\,=\,\,v^*\,-\,\frac{\pi^2D\,\cosh\lambda^*}{\ln^2N}\,.
\label{e7}
\end{eqnarray}

The essential correctness of Eq. (\ref{e7}), that the convergence of
the asymptotic front speed to $v^*$ behaves $\sim1/\ln^2N$ has been
observed elsewhere in systems of physical interest such as clock model
\cite{ramses}, in field-theoretical approach \cite{levine}, and as
well as in studies on noisy front propagation in Fisher equation
in the limit of ``weak noise'' by mathematicians \cite{doering}.
However, it has also been noted by many theorists working in this
field that one often has to go to very high values of $N$ to obtain
the $1/\ln^2 N$ convergence of the asymptotic front speed to $v^*$.
For large but more reasonable values of $N$, there are often
significant deviations from Eq. (\ref{e7}) (see for example, Ref.
\cite{kns}). A question which then naturally arises is the following:
how does one bridge the gap between Eq. (\ref{e7}), which is valid for
asymptotically large $N$, and the corresponding results for finite
$N$?

Based on the discussion three paragraphs above, in what follows below,
we will appeal to the reader for a more comprehensive description of
fluctuating ``pulled'' fronts on a lattice that combines the
stochastic halt-and-go nature of front propagation at the f.o.l.s.
with a uniformly translating front solution few lattice
sites behind the f.o.l.s. Such a description leads one to a concrete
mathematical formalism developed elsewhere \cite{pvs}, of which we
will provide only the flavour in this paper. This description and the
associated formalism serves to bridge this gap between the results for
asymptotically large $N$ and that for reasonably large values of
$N$. Afterwards, we will argue why Eq. (\ref{e7}) is essentially
correct for asymptotically large $N$ values. The mathematical
formalism of Ref. \cite{pvs}, however, has open questions, which we
will leave out for the next section.

We start with the fact that the very definition of the f.o.l.s. means
that all the lattice sites on the right of it are empty (and they have
never been occupied before). Naturally, a lattice site, which has
never been occupied before attains the status of the f.o.l.s. as soon
as one  particle hops into it from the left. In reference to the
lattice, the position of the f.o.l.s. remains fixed at this site for
some time, i.e., after its creation, an f.o.l.s.  remains the
f.o.l.s. for some time. During this time, however, the number of
particles on and behind the f.o.l.s. continues to grow. As the number
of particles grows on the f.o.l.s., the chance of one of them making a
diffusive hop on to the right also increases. At some instant, a
particle from the f.o.l.s. hops over to the right: as a result of this
hop, the position of the f.o.l.s.  advances by one unit on the
lattice, or, viewed from another angle, a new f.o.l.s. is created
which is one lattice distance away on the right of the previous
one. Microscopically, the selection process for the length of the time
span between two consecutive f.o.l.s. creations is stochastic, and the
inverse of the long time average of this time span defines the front
speed. Simultaneously, the amount of growth of particle numbers on and
behind the f.o.l.s. itself depends on the time span between two
consecutive f.o.l.s. creations (the longer the time span, the longer
the amount of growth). As a consequence, on average, the selection
mechanism for the length of the time span between two consecutive
f.o.l.s. movements, which determines the asymptotic front speed, is
nonlinear.

This inherent nonlinearity makes the prediction of the asymptotic
front speed difficult. One might recall the difficulties associated
with the prediction of pushed fronts due to nonlinear terms in this
context, although the nature of the nonlinearities in these two cases
is {\it completely different\/}. In the case of pushed fronts, the
asymptotic front speed is determined by the mean-field dynamics of the
fronts, and the nonlinearities originate from the {\it nonlinear growth
terms in the partial differential equations\/} that describe the
mean-field dynamics. On the other hand, for fronts consisting of
discrete particles on a discrete lattice, the corresponding mean-field
growth terms are {\it linear\/}, but since the asymptotic front speed
is determined from the probability distribution of the time span
between two consecutive f.o.l.s. movements, on average, it is the
relation between this probability distribution and the effect of the
linear growth terms that the nonlinearities stem from. As one can now
clearly see, Brunet and Derrida's cutoff picture, as described above,
does not take the stochasticity of the halt-and-go mechanism into
account.

Our approach is to discuss a separate probabilistic theory for the
hops to create the new f.o.l.s., and then to demonstrate that by
matching the description of the behaviour in this region to the more
standard one (of growth and roughly speaking, uniform translation)
behind it, one obtains a consistent and more complete description of
the stochastic and discreteness effects on front propagation. In the
simplest approximation, the theory provides a very good fit to the
data, but our approach can be systematically improved by incorporating
the effect of fluctuations as well \cite{pvs}. Besides providing
insight into how a stochastic front propagates at the far tip of the
leading edge, our analysis naturally leads to a more complete
description that allows one to  interpret the finite $N$ corrections
to the front speed for much smaller values of $N$ than that are
necessary to see the asymptotic result of Brunet and Derrida
\cite{bd}. We stress here that such a procedure can be carried out for
any fluctuating ``pulled'' front, although in the present context, we
will confine ourselves only to the reaction-diffusion system X
$\leftrightarrow$ 2X.

In the resulting mathematical formalism \cite{pvs}, we follow the
movement of the f.o.l.s. of {\it one single front realization\/} over
a long time at large times. Let us denote the $j$ successive values of
the duration of halts of the f.o.l.s. by $\Delta t_1, \Delta
t_2,\ldots,\Delta t_j$. One can then define the front speed as
\begin{eqnarray}
v_N\,=\,\lim_{j\rightarrow\infty}\,j\,\left[\sum_{j'=1}^{j}\Delta
t_{j'}\right]^{-1}\,.
\label{e8}
\end{eqnarray}
Put in a different way, if we denote the probability that a
f.o.l.s. remains the f.o.l.s. for time $\Delta t$ by ${\cal P}(\Delta
t)$, the asymptotic front speed, according to Eq. (\ref{e8}), is given
by
\begin{eqnarray} 
v_N\,=\,\left[\,\int_0^\infty d(\Delta t)\,\,\Delta t\,\,{\cal
P}(\Delta t)\,\right]^{-1}\,.
\label{e9}
\end{eqnarray}
\begin{figure}
%\begin{center}
\includegraphics[width=0.75\linewidth]{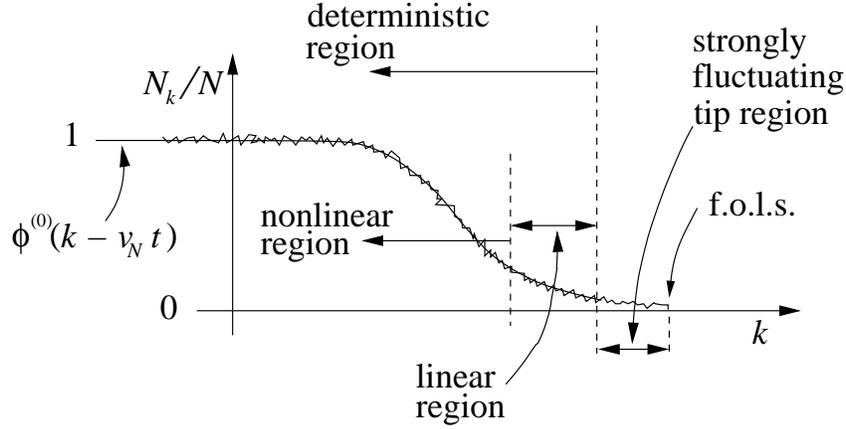}
\caption{A snapshot of a realization of the reaction-diffusion system
X $\leftrightarrow$ 2X (shown by the jaggered line) as
$t\rightarrow\infty$, and how a theorist would picture such a
front. In this picture, as the smooth line shows, a uniformly
translating solution travelling with speed $v_N$, given by
$\phi_k(t)=\phi^{(0)}(k-v_Nt)$ and obeying Eq. (\ref{e5}), is suitable
(in an average sense) everywhere but few lattice sites at the tip of
the front leading up to the f.o.l.s. The region where such a
description holds is denoted above by ``deterministic region''. The
``deterministic region'' is further subdivided into two parts --- in
the ``linear region'', the nonlinear term
$\displaystyle{\left[\phi^{(0)}\right]^2}$ of Eq. (\ref{e5}) can be
neglected. In the ``nonlinear region'', however, all the terms of
Eq. (\ref{e5}) have to be taken into account.\label{fig5}\vspace{-4mm}}
%\end{center}
\end{figure}
Our goal is to obtain a theoretical expression of ${\cal P}(\Delta t)$,
but to do so, we draw the reader's attention to Fig. \ref{fig5}. As
described above, we assume that a uniformly translating (with speed
$v_N$) deterministic mean-field description, given by
$\phi_k(t)=\phi^{(0)}(k-v_Nt)$, is valid everywhere {\it but a few
lattice sites leading up to the f.o.l.s.\/}. The region where such a
description [namely that its dynamics is given by Eq. (\ref{e5})]
holds is denoted by ``deterministic region'' in Fig. \ref{fig5}. This
region can be further subdivided into two parts, a ``linear region''
and a ``nonlinear region'', where the nonlinear
$\displaystyle{\left[\phi^{(0)}\right]^2}$ term can and cannot be
neglected respectively. Although the amount of fluctuations in the
number of particles on individual lattice sites at the strongly
fluctuating tip region is of the same order of magnitude as the
particle numbers themselves, following Ref. \cite{pvs}, we assume that
the strongly fluctuating tip region can be modelled by a time
dependent mean-field type description without uniform translation. In
this description, at the tip region of the front, we express the front
solution as $\phi_k(t)=\phi^{(0)}(k-v_Nt)+\delta\phi_k(t)$. The
quantities $\delta\phi_k(t)$ are non-zero due to the halt-and-go
motion of the f.o.l.s., but at the boundary between the tip region and
the ``linear region'' $\delta\phi_k(t)$ vanishes.  As explained
previously, to obtain the expression of ${\cal P}(\Delta t)$, one
needs the expressions of $\phi^{(0)}(k-v_Nt)$ and
$\delta\phi_k(t)$. Moreover, the expression of $v_N$ itself is needed
to solve for  $\phi^{(0)}(k-v_Nt)$, and $v_N$ can be determined only
from ${\cal P}(\Delta t)$ as Eq. (\ref{e9}) shows. This indicates that
the only way to obtain the expression of ${\cal P}(\Delta t)$ is to
solve a whole system of equations self-consistently \cite{pvs}. We
also note here that in this self-consistent theory, there is an
effective parameter.

\begin{figure}
%\begin{center}
\vspace{-2mm}
\includegraphics[width=0.9\linewidth]{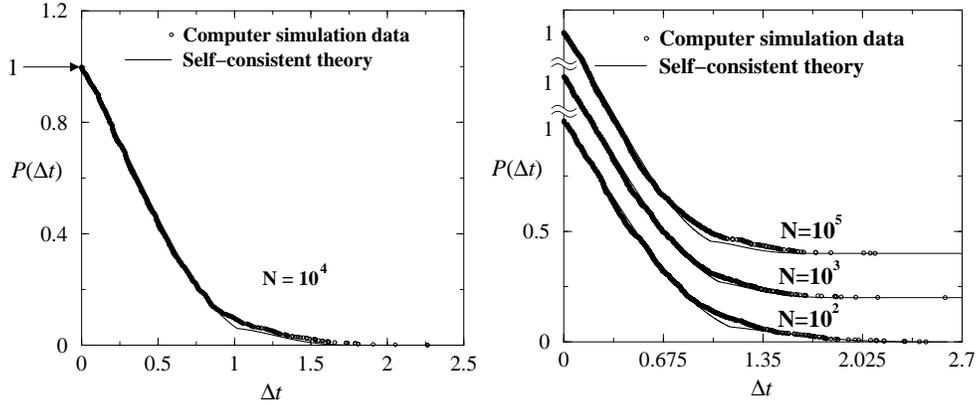}
\caption{The agreement between the self-consistent theoretical curve
of $P(\Delta t)$ {\protect\cite{pvs}} and computer simulation
results.\label{fig6}\vspace{-4mm}}
%\end{center}
\end{figure}

This self-consistent set of equations are highly nonlinear and
complicated, but due to the presence of the effective parameter in our
theory, our procedure to obtain the ${\cal P}(\Delta t)$ is {\it not
predictive\/}. However, the fact that it the theory generates a
probability distribution that agrees so well with numerical
simulations is indicative of the essential correctness of such a
description \cite{pvs} of a fluctuating ``pulled'' front. The
self-consistent theoretical curves of $P(\Delta
t)=\displaystyle{\int_{\Delta t}^\infty dt'\,{\cal P}(t')}$ for $D=1$
and $N=10^4, 10^2, 10^3$ and $10^5$ (in that order) are shown in
Fig. \ref{fig6}. The corresponding numerical comparison of front
speeds are shown in Table I. First, we notice that in the graphs of
Fig. \ref{fig6}, the theoretical curves lie below the simulation
histograms at $\Delta t\approx2/v_N$ --- this is an artifact of the
matching that we had to do for the expressions of $P(\Delta t)$ below
and above $\Delta t\sim2/v_N$. This difference occurs due to certain
fluctuation and correlation effects \cite{pvs}. Secondly, the
agreement between the simulation data and the self-consistent theory
is not very good for $N=10^5$ --- at this value of $N$, the simulation
gets very slow and one has to continuously remove particles from the
saturation region of the front to gain program speed, which affects
the $P(\Delta t)$ histograms significantly for large times.

\begin{center}
\begin{tabular}{||@{}c|c@{}|c@{}|c@{}||}
\hline\hline
\rule[-2mm]{0pt}{4ex}\,\,\,\,$N$\,\,\,\,&\,\,\,\,$v_N\mbox{(simulation)}$
\,\,\,\,&\,\,\,\,$v_N\mbox{(theoretical)}$\,\,\,\,&\,\,\,\,$v_N\mbox{[Eq.
(\ref{e7})]}$\,\,\,\,\\  \hline
\rule[-2mm]{0pt}{4ex}$10^2$&\,\,\,\,$1.778$\,\,\,\,&
\,\,\,\,$1.808$\,\,\,\,&$1.465$\\ \hline
\rule[-2mm]{0pt}{4ex}$10^3$&\,\,\,\,$1.901$\,\,\,\,&
\,\,\,\,$1.899$\,\,\,\,&$1.803$\\ \hline
\rule[-2mm]{0pt}{4ex}$10^4$&\,\,\,\,$1.964$\,\,\,\,&
\,\,\,\,$1.988$\,\,\,\,&$1.925$\\ \hline
\rule[-2mm]{0pt}{4ex}$10^5$&\,\,\,\,$2.001$\,\,\,\,&
\,\,\,\,$2.057$\,\,\,\,&$1.976$\\ \hline\hline
\end{tabular}
\end{center}

{\footnotesize Table I: Comparison of $v_N$ values, simulation, self
consistent theory \cite{pvs} [indicated by $v_N$(theoretical)], and
that of Eq. (\ref{e7}).}

In addition to such good agreements between our self-consistent theory
and simulations for the $P(\Delta t)$ curves, a significant
observation is the following: as the value of $N$ is increased in our
self-consistent theory, we find that the quantity
$\delta\phi_k/\phi^{(0)}_k$ values at the tip of the front gradually
reduces \cite{pvs}. The stochastic halt-and-go character of the
movement of the f.o.l.s., which is usually occupied by ${\cal O}(1)$
number of particles, however, continues to remain valid for any value
of $N$. This implies that for very large $N$, one approaches the
picture of a fluctuating ``pulled'' front, where a uniformly
translating mean-field description (\ref{e5}) holds all the way up to
one lattice site behind the f.o.l.s., while only the dynamics of the
f.o.l.s. is a stochastic halt-and-go process. Such a simplified
picture has been studied numerically in the last paper of
Ref. \cite{bd}.

We finally end this section with a short note arguing why the
expression (\ref{e7}) is correct for asymptotically large $N$. In
fact, it can be understood very simply when one observes that the
length of the ``linear region'' in Fig. \ref{fig5} increases as $\ln
N$, and combines it with the expectation that the asymptotic front
speed should be $<v^*$ (as motivated in the origin of the terminology
``pulled'' front). For a deterministic pulled front equation, such as
Eqs. (\ref{e1}) or (\ref{e5}), the length of the linear region is
infinite, and in such cases, in the absence of any length scale, the
stability criterion of the front solution in the comoving frame
dictates that for the selected front speed, it is good enough to
consider exponentially decaying solution
$\phi(\xi)\sim\exp[-\lambda\xi]$ with
$\lambda\equiv\mbox{Re}(\lambda)$. On the other hand, for ``pulled
fronts'' consisted of discrete particles on a lattice, the positivity
of $\phi(\xi)$ demands that one has to consider an oscillatory
function of wavelength $\pi/\ln N$ as a multiplicative factor to the
exponentially decaying $\exp[-\lambda\xi]$ profile for the front
solution $\phi(\xi)$ to obtain an asymptotic speed $<v^*$. Just from
this argument alone, it is possible to derive the asymptotic front
speed of Eq. (\ref{e7}).

\section{Outlook and Unsolved Problems}

When an equation that allows so-called pulled fronts in the mean-field
limit is modelled with a stochastic model with a finite number $N$ of
particles per correlation volume, the convergence to the speed $v^*$
for asymptotically large $N$, as obtained by Brunet and Derrida,
behaves as $\ln^{-2}N$, and this behaviour is model
independent. However, for large but more reasonable values of $N$,
there are significant deviations from their result, and these
deviations stem from the complicated stochastic halt-and-go dynamics
of the foremost occupied lattice site, where the actual microscopic
rules of the system under consideration play a crucial role, and
therefore the deviations from the results of Brunet and Derrida are
model dependent.

The message of this paper is as follows: to obtain the deviations from
the $1/\ln^2N$ convergence to $v^*$ one really has to focus at the
stochastic halt-and-go movement of the foremost occupied lattice
site. From the mean-field limit of this fluctuating front, we know
that the tip region is very important for its dynamics; as a result,
the fluctuating tip plays a very significant role in deciding the
asymptotic front speed, in which two very important aspects come to
play a role --- discrete nature of particles and discrete nature of
the lattice indices. In this paper, we have outlined the formulation
of a self-consistent theory, which we developed in Ref. \cite{pvs}, to
model the fluctuating tip by means of a mean-field type
description. The mean-field type description of the fluctuating tip is
then matched to a uniformly translating solution behind. Due to the
presence of an effective parameter in this self-consistent theory, it
is {\it not predictive\/}. How to obtain a predictive theory for
moderately large values of $N$, and demonstrate analytically how the
corresponding front speed approaches to the expression (\ref{e7})
still remain unsolved problems. However, one has to remember that in
actuality, the tip is strongly fluctuating and there are
time-correlation effects involved (for a more detailed discussion, see
Sec. IV.C of Ref. \cite{pvs}). Any alternative predictive theory, that
one might think at this point, must be able to take these into
account, in addition to the mean-field type self-consistent theory of
Ref. \cite{pvs}.

The prospect of such a theory however, looks grim at this point. Not
only the problem becomes highly nonlinear, but also one must realize
that the fluctuations in the number of X particles on the lattice
sites at the tip of the front is of the same order as the number of X
particles in them ($\sim1$), and there does not exist any small
parameter that one can do perturbation theory with.


\begin{thebibliography}{99}

%1
\bibitem{meron} E. Meron, Phys. Rep. {\bf 218}, 1 (1992).

%2
\bibitem{tabeling} G. Zocchi, P. Tabeling and M. Ben Amar,
Phys. Rev. Lett. {\bf 69}, 601 (1992).

%3
\bibitem{tucross} Y. Tu and M. C. Cross, Phys. Rev. E {\bf 69}, 2515
(1992).

%4
\bibitem{streamers} U. Ebert, W. van Saarloos and C. Caroli,
Phys. Rev. E {\bf 55}, 1530 (1997).

%5
\bibitem{ch} M. C. Cross and P. C. Hohenberg, Rev. Mod. Phys. {\bf
65}, 851 (1993).

%6
\bibitem{bj} E. Ben-Jacob, H. R. Brand, G. Dee, L. Kramer,  and
J. S. Langer, Physica D {\bf 14}, 348 (1985).

%7
\bibitem{vs2} W. van Saarloos, Phys. Rev. A {\bf 39},  6367 (1989).

%8
\bibitem{ebert} U. Ebert and W.  van Saarloos, Physica D {\bf 146}, 1
(2000).

%9
\bibitem{fisher} R. A. Fisher, Ann. Eugenics {\bf 7}, 355 (1937).

%10
\bibitem{bramson} M. Bramson, Mem. Am. Math. Soc. {\bf 44}, No. 285
(1983).

%11
\bibitem{breuer} H. P. Breuer, W. Huber, and F. Petruccione, Physica D
{\bf 73}, 259 (1994).

%12
\bibitem{lemar} A. Lemarchand, A. Lesne, and M. Mareschal,
Phys. Rev. E {\bf 51}, 4457 (1995).
 
%13
\bibitem{armeroprl} J. Armero, J. M. Sancho, J. Casademunt,
A. M. Lacasta, L. Ram\'{\i}rez-Piscina, F. Sagu\'es,
Phys. Rev. Lett. {\bf 76}, 3045 (1996).

%14
\bibitem{bd} E. Brunet and B. Derrida, Phys. Rev. E {\bf 56}, 2597
(1997). E. Brunet and B. Derrida, Comp. Phys. Comm. {\bf 122}, 376
(1999). E. Brunet and B. Derrida, J. Stat. Phys., {\bf 103}, 269 (2001).

%15
\bibitem{kns} D. A. Kessler, Z. Ner, and L. M. Sander,
Phys. Rev. E. {\bf 58}, 107 (1998). 

%16
\bibitem{levine} L. Pechenik and  H. Levine, Phys. Rev. E {\bf 59},
3893 (1999).

%17
\bibitem{armero} J. Armero, J. Casademunt, L. Ram\'{\i}rez-Piscina,
and J.M. Sancho, Phys. Rev. E {\bf 58}, 5494 (1998).

%18
\bibitem{riordan}  J. Riordan, C. R. Doering, and D. ben-Avraham,
Phys. Rev. Lett. {\bf 75}, 565 (1995).

%19
\bibitem{dee} G. Dee and J. S. Langer, Phys. Rev. Lett. {\bf 50}, 383
(1983).

%20
\bibitem{ramses} R. van Zon, H. van Beijeren and Ch. Dellago,
Phys. Rev. Lett. {\bf 80}, 2035 (1998).

%21
\bibitem{pvs2} D. Panja and W. van Saarloos, Phys. Rev. E {\bf 66},
015206(R) (2002).

%22
\bibitem{rocco} A. Rocco, U. Ebert, W. van Saarloos,
Phys. Rev. E. {\bf 62}, R13 (2000).

%23
\bibitem{tripathy1} G. Tripathy and W. van Saarloos,
Phys. Rev. Lett. {\bf 85}, 3556 (2000).

%24
\bibitem{tripathy2} G. Tripathy, A. Rocco, J. Casademunt, and W. van
Saarloos, Phys. Rev. Lett. {\bf 86}, 5215 (2001).

%25
\bibitem{moro} E. Moro, Phys. Rev. Lett. {\bf 87}, 238303 (2001).

%26
\bibitem{pvs} D. Panja and W. van Saarloos, Phys. Rev. E {\bf 66}, 
036206 (2002).

%27
\bibitem{pvs3} D. Panja and W. van Saarloos; Phys. Rev. E {\bf 65},
057202 (2002).  

%28
\bibitem{doering} C. R. Doering, C. Mueller, and P. Smereka, {\it 
"Noisy Wavefront Propagation in the Fisher-Kolmogorov-Petrovsky-Piscunov 
Equation"\/}, UPoN 2002 (this volume); {\it 
"Interacting Particles, the Stochastic Fisher-Kolmogorov-Petrovsky-Piscunov 
Equation, and Duality"\/}, Physica A, submitted (2002).

\end{thebibliography}
\end{document}